\documentclass[conference, review]{IEEEtran}
\usepackage{siunitx}
\usepackage{amsmath,amssymb,amsfonts, mathtools}
\usepackage{multirow}
\usepackage{algorithm,algorithmic}
\usepackage{ragged2e}
\usepackage[hyphens]{url}
\usepackage[draft]{hyperref}
\usepackage{cite}
\usepackage{graphicx}
\graphicspath{{./fig/}}
\DeclareGraphicsExtensions{.pdf,.jpeg,.png}
\usepackage[caption=false,font=footnotesize]{subfig}
\usepackage{array}

\newcommand{\myTitleFirstPage}[0]{Low-Rank Compression for IMC Arrays}

\newcommand{\myAuthors}[0]
    {
        Kang Eun Jeon\textsuperscript{1}, 
        Johnny Rhe\textsuperscript{1} and 
        Jong Hwan Ko\textsuperscript{1,2}
    }

\newcommand{\myEmails}[0]
    {
        Email: \{kejeon, 
        djwhsdj, 
        jhko\}@skku.edu
    }

\newcommand{\myAuthorBlock}[0]{
    \author{
        \IEEEauthorblockN{\myAuthors}
        \IEEEauthorblockA{
        \textsuperscript{1}Department of Electrical and Computer Engineering, Sungkyunkwan University, Suwon, Korea \\
        \textsuperscript{2}College of Information and Communication Engineering, Sungkyunkwan University, Suwon, Korea\\ 
        \myEmails
            }
        }
    }

\newcommand{\mysubsection}[1]{\vspace{0.5em}\noindent\textbf{#1}}

\newcommand{\mysubsectionNospace}[1]{\noindent\textbf{#1}}
    
\begin{document}
\title{\myTitleFirstPage}

\hyphenation{ConvMapSIM}
\myAuthorBlock

\maketitle

\begin{abstract}
In this study, we address the challenge of low-rank model compression in the context of in-memory computing (IMC) architectures.
Traditional pruning approaches, while effective in model size reduction, necessitate additional peripheral circuitry to manage complex dataflows and mitigate dislocation issues, leading to increased area and energy overheads.
To circumvent these drawbacks, we propose leveraging low-rank compression techniques, which, unlike pruning, streamline the dataflow and seamlessly integrate with IMC architectures. However, low-rank compression presents its own set of challenges, namely i) suboptimal IMC array utilization and ii) compromised accuracy. To address these issues, we introduce a novel approach i) employing shift and duplicate kernel (SDK) mapping technique, which exploits idle IMC columns for parallel processing, and ii) group low-rank convolution, which mitigates the information imbalance in the decomposed matrices. Our experimental results
demonstrate that our proposed method 
achieves up to 2.5$\times$ speedup or +20.9\% accuracy boost over existing pruning techniques.
\end{abstract}

\section{Introduction}
The advent of in-memory computing (IMC) architecture heralds a transformative shift in the computing domain, primarily driven by the escalating demands for processing large-scale data on complex deep neural networks. 
By unifying computation and data storage, IMC overcomes the von Neumann bottleneck inherent in traditional architectures that separate memory and processing units. 
This integration facilitates direct matrix-vector multiplication (MVM) within the memory itself, exploiting the parallel computation capabilities for expedited processing at lower energy costs \cite{pipelayer}. 

Despite these advantages, IMC architectures face challenges in handling convolution operations, which require reshaping of convolutional weights and input data for MVM compatibility. The image-to-column (im2col) method \cite{im2col} unrolls convolutional weights into IMC columns for MVM but often suffers from low column utilization.
To address this, techniques such as shift and duplicate kernel (SDK) \cite{sdk} and variable-window SDK (VW-SDK) \cite{vwc-sdk} have been proposed. These methods enhance array utilization and computational performance by enabling input data reuse and parallel processing, effectively exploiting idle columns where duplicated kernels are situated.


\begin{figure}
    \centering
    \includegraphics[width=0.9\columnwidth]{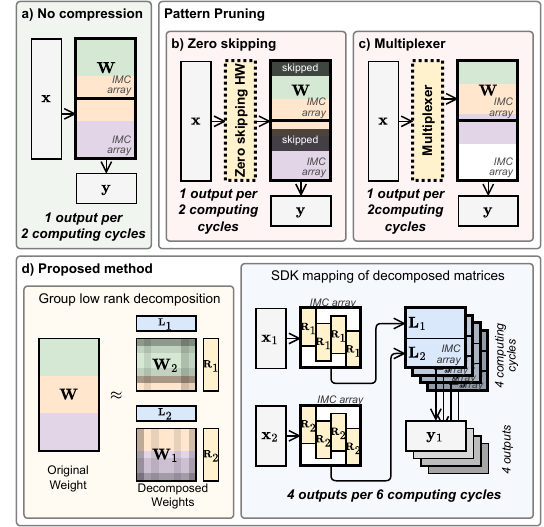}
    \caption{Conventional model compression methods for IMC arrays and the proposed low-rank compression method. }
    \vspace{-1em}
    \label{fig:paper-overview}
\end{figure}

While mapping techniques \cite{im2col, sdk, vwc-sdk} improve array utilization, they do not compress the weights themselves for additional performance gains. Pruning methods \cite{vwc-sdk, patdnn, pairs, exploring, unst_pruning}, particularly structured pruning \cite{pairs, exploring} tailored to the unique hardware constraints of IMC arrays, emerged as a promising solution. Structured pruning reduces the computational workload by omitting non-essential weights in a way that complements the IMC's MVM functionality. 

However, pruning techniques encounter hurdles due to the necessity of additional peripheral circuitry, such as zero-skipping hardware \cite{zero-skipping, islped_row_merging} or multiplexers \cite{exploring}, to translate model sparsity into performance benefits (see Fig. \ref{fig:paper-overview}). Zero-skipping hardware leverages sparsity by deactivating unnecessary wordlines—rows containing zero-valued weights—while multiplexers realign input data with pruned weights to counteract dislocation. These requirements introduce extra area and energy overheads, hindering the practical adoption of pruning methods despite their theoretical advantages.


To overcome the drawbacks associated with pruning, this research advocates for low-rank matrix decomposition technique as an alternative for compressing neural network weights. Unlike pruning, low-rank compression does not necessitate complex peripheral circuitry or realignment mechanisms, offering a more straightforward integration into IMC arrays. However, this method typically involves a trade-off between compression and accuracy, with low-rank compression often resulting in lower performance compared to pruning. Moreover, low-rank compressed matrices frequently lead to suboptimal utilization of IMC arrays. Our work, as shown in Fig. \ref{fig:paper-overview}d, introduces new techniques, namely, SDK and group low-rank compression, aiming to balance accuracy retention, and IMC array utilization. 
Our experimental results show that our proposed method can achieve up to 2.5$\times$ speedup and +20.9\% accuracy boost on Wide ResNet16-4 versus pruning methods. 

\section{Backgrounds and Related Works}
\begin{figure}
    \centering
    \includegraphics[width=0.9\columnwidth]{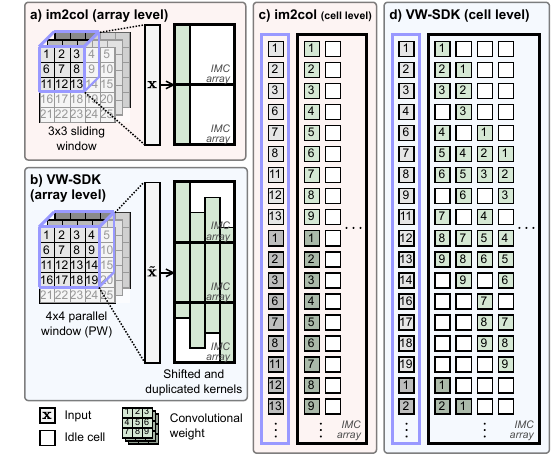}
    \vspace{-1em}
    \caption{Convolutional weight mapping methods.}
    \vspace{-1em}
    \label{fig:conv_weight_mapping}
\end{figure}

\mysubsectionNospace{IMC and Convolutional Weight Mapping.} 
IMC architecture marks a paradigm shift towards memory-centric computing, where MVM operation is performed directly within the memory that hosts the deep learning model parameters. 
While IMC architecture is adept at MVM operations, it is not inherently equipped for convolution operations. To address this, convolutional weight mapping methods such as image to column (im2col) have been employed. Im2col, as illustrated in Fig. \ref{fig:conv_weight_mapping}a and c, maps a sliding window of the input feature map (IFM) to the input port of the IMC array. Concurrently, it unrolls and maps each output channel of the kernel to the columns of the IMC array, thus facilitating the convolution operation in the form of MVM.
However, since the number of utilized columns equals the number of output channels, the array utilization of im2col mapping is contingent upon the number of output channels. Hence, im2col mapping often delivers suboptimal array utilization with smaller convolutional filters and consequently resulting in additional computing cycles.

To address the low array utilization issue of im2col \cite{im2col}, Zhang et al. \cite{sdk} and Rhe et al. \cite{vwc-sdk} proposed shift and duplicate kernel (SDK) mapping method. The SDK method uses parallel window (PW) and duplicated kernels to facilitate parallel processing of multiple sliding windows concurrently—unlike the single window processing inherent to the im2col method. By situating duplicated kernels in previously idle columns of the IMC array, the SDK method significantly enhances array utilization. The extent of this enhancement is governed by the size of the PW; for instance, employing a 4$\times$4 PW allows for the duplication of three additional kernels, thereby increasing the number of simultaneously processed sliding windows. 
However, as illustrated in Fig. \ref{fig:conv_weight_mapping}b and d, SDK mapping introduces structural sparsity by its very nature. Specifically, larger PW sizes improve idle column utilization, but at the expense of increased sparsity within the rows. 
Based on the generated mapping, the computing cycle of IMC array can be calculated, as proposed by Rhe et al. \cite{vwc-sdk}, using array row (AR) and array column (AC) cycles. AR cycle defines the number of arrays required to process the rows in a mapping, and AC cycle the columns.

\mysubsection{Pruning Methods and Challenges on IMC.} Weight pruning technique \cite{unst_pruning, vwc-sdk} is a strategy to reduce computational requirements by eliminating redundant or non-contributory weights within neural networks.
Within the IMC community, recent pruning techniques are tailored to compress the weight matrix to fit the constraint of IMC arrays, thereby enhancing computational efficiency.
For example, Rhe et al. \cite{vwc-sdk} have proposed the column-wise pruning method to exploit the structural column sparsity of the weight matrix through channel pruning, achieving 1.38$\times$ inference speed in ResNet-20.
Similarly, pattern-based pruning has been used to compress the weight matrix in the row direction, which shows up to 4$\times$ higher compression rate \cite{pairs}.
Although these pruning techniques have shown promising results in compressing the weight matrix and improving computation performance on IMC arrays, these methods necessitate additional peripheral circuits, such as Multiplexer (MUX) \cite{exploring} and Demultiplexer (DEMUX) \cite{structured}, which aims to remap the data path of the input feature for realignment with the sparsity pattern of the pruned model \cite{pimprune, flexible}.
Consequently, while these pruning methods have been instrumental in enhancing the computational efficiency of IMC arrays, the necessity for supplementary peripheral circuits impedes their real-life adoption.


\mysubsection{Low-Rank Compression.} 
To exploit the low-rank properties inherent in neural network weights, various low-rank compression techniques have been applied with considerable success \cite{denton_low-rank}. 
While the low-rank compression method is often considered to be less effective compared to other compression methods, it holds a significant advantage in terms of fast inference, especially on GPUs \cite{trp_low-rank}. This is attributed to its use of dense matrices, which exhibit local, regular, and parallelizable memory access patterns, facilitating quicker computations. 
The previous research efforts \cite{yani_low-rank,trp_low-rank}, while pioneering in advancing low-rank compression techniques for deep neural networks, are mostly tailored for optimization on GPUs.
However, this focus has inadvertently left a gap in the exploration of low-rank compression techniques for other forms of hardware, particularly IMC arrays. IMC arrays, known for their potential to significantly reduce energy consumption and latency in performing matrix operations, present a unique architecture that could benefit from specialized compression methods. Yet, the application of low-rank compression within the context of IMC arrays remains unexplored, signifying a critical research gap. This oversight underscores the need for a dedicated investigation into how low-rank compression techniques can be adapted or reimagined to exploit the distinctive advantages and architecture of IMC arrays, a challenge that our current research endeavors to address.

\begin{figure}
    \centering
    \includegraphics[width=0.8\columnwidth]{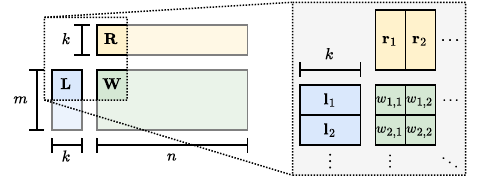}
    \vspace{-1em}
    \caption{Low-rank matrix decomposition.}
    \vspace{-1em}
    \label{fig:low-rank_decomposition}
\end{figure}



\section{Motivation}
\begin{figure}
    \centering
    \includegraphics[width=0.9\columnwidth]{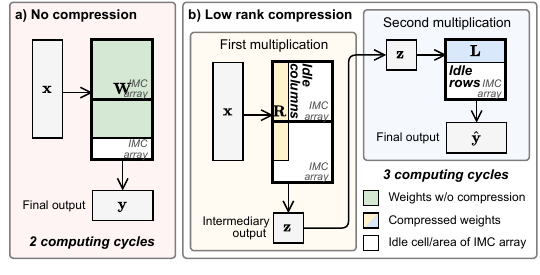}
    \vspace{-1em}
    \caption{Motivation of our research.}
    \vspace{-1em}
    \label{fig:challenges}
\end{figure}

Given a weight matrix $\mathbf{W} \in \mathbb{R}^{m \times n}$, low-rank decomposition approximates it as $\hat{\mathbf{W}} = \mathbf{L}\mathbf{R}$, where $\mathbf{L} \in \mathbb{R}^{m \times k}$ and $\mathbf{R} \in \mathbb{R}^{k \times n}$. Each element $w_{i,j}$ is the dot product of the $i^\text{th}$ row of $\mathbf{L}$ and the $j^\text{th}$ column of $\mathbf{R}$ (see Fig. \ref{fig:low-rank_decomposition}). The parameter $k$ balances approximation accuracy and computational savings; smaller $k$ means more compression but potentially more information loss.
The adaptation of the low-rank compression technique, which involves decomposing a larger matrix into two smaller ones, to IMC architecture presents two significant impediments: low array utilization and diminished accuracy in machine learning tasks. Fig. \ref{fig:challenges} illustrates the computational difficulties of applying low-rank matrix compression within an IMC framework. For instance, when original, uncompressed convolutional weights are mapped onto IMC arrays using the prevalent im2col mapping strategy, a rectangular-shaped weight matrix, $\mathbf{W}$, is produced. This matrix extends across more rows than columns and requires three computing cycles to generate a single output, as shown in Fig. \ref{fig:challenges}a. Conversely, Fig.  \ref{fig:challenges}b showcases low-rank compression on IMC arrays, where $\mathbf{W}$ is decomposed into two matrices that do not fully utilize the IMC array's capacity. This decomposition, intended to reduce computational load, paradoxically introduces an additional computing cycle due to low array utilization.

Moreover, the inherent rectangular shape of convolutional kernels leads to a significant imbalance in information encoding between the $\mathbf{L}$ and $\mathbf{R}$ matrices. This imbalance causes a notable loss of information in the weight matrix's rows, thereby reducing computation accuracy, a crucial aspect for the effectiveness of neural network models. To address the first challenge of low array utilization, we propose the integration of SDK mapping with the low-rank compression technique. This approach enhances array utilization through input data reuse and the added parallelism of duplicated kernels. For the second challenge, concerning reduced machine learning task accuracy, we introduce grouped low-rank decomposition. This method partitions the weight matrix into multiple groups prior to low-rank compression, effectively mitigating the information imbalance initially present in $\mathbf{L}$, while capturing essential weight features with a minimal increase in parameters.

\begin{figure*}
    \centering
    \includegraphics[width=1.7\columnwidth]{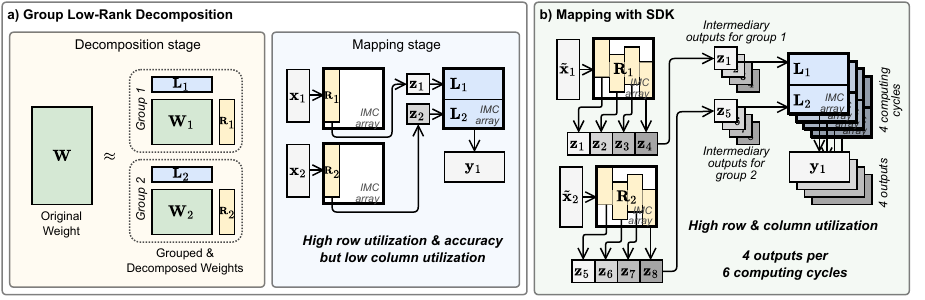}
    \vspace{-1em}
    \caption{Overview of the proposed techniques for low-rank compression on IMC arrays.}
    \vspace{-1em}
    \label{fig:proposed}
\end{figure*}

\section{Proposed Method}
\mysubsectionNospace{Group Low-Rank Compression.} 
To address the severe accuracy degradation and low row utilization issues associated with the $\mathbf{L}$ matrix in traditional low-rank approximations, we propose a group low-rank decomposition technique, as illustrated in Fig.~\ref{fig:proposed}a. 
In this approach, the weight matrix is partitioned into $g$ submatrices or groups, denoted in block matrix notation as $\mathbf{W} = [\mathbf{W}_1, \mathbf{W}_2, \dots, \mathbf{W}_g]$, where $g$ is the number of groups. Each submatrix $\mathbf{W}_i$ is then independently compressed using low-rank decomposition:
\begin{equation}
    \mathcal{D}_g(\mathbf{W}) := \begin{bmatrix}
        \mathcal{D}(\mathbf{W}_1),\;
        \mathcal{D}(\mathbf{W}_2),\; 
        \cdots,\;
        \mathcal{D}(\mathbf{W}_g)
    \end{bmatrix}.
\end{equation}
Here, $\mathcal{D}_g(\cdot)$ denotes the grouped low-rank decomposition operator for a specified number of groups $g$, and $\mathcal{D}(\cdot)$ represents the traditional low-rank decomposition operator without matrix partitioning, such that $\mathcal{D}(\mathbf{W}_i) := \mathbf{L}_i \mathbf{R}_i$.

\vspace{1em}\noindent\textbf{Theorem 1.} \textit{Given a weight matrix $\mathbf{W}$ and a target rank $k$, the reconstruction error of its group low-rank approximation, $\varepsilon_g := ||\mathbf{W} - \mathcal{D}_g(\mathbf{W})||_\mathrm{F}$, is upper-bounded by that of the traditional low-rank approximation, $\varepsilon := ||\mathbf{W} - \mathcal{D}(\mathbf{W})||_\mathrm{F}$, for an arbitrary number of groups, $g$:} 
\begin{equation}
    \underbrace{||\mathbf{W} - \mathcal{D}_g(\mathbf{W})||_\mathrm{F} }_{\varepsilon_g}
    \leq 
    \underbrace{||\mathbf{W} - \mathcal{D}(\mathbf{W})||_\mathrm{F}}_{\varepsilon}
    \label{eqn:theorem1}
\end{equation}
\textit{where both reconstruction errors are measured in Frobenius norm, denoted by $||\cdot||_\mathrm{F}$.}

\vspace{1em}\noindent\textit{Proof.} We begin by approximating $\mathbf{W}$ using truncated singular value decomposition (SVD), i.e., $\mathcal{D}(\mathbf{W}) = \mathbf{U}\mathbf{\Sigma}\mathbf{V}^\top$. We know that this is an optimal approximation with respect to the Frobenius norm according to the Eckart-Young-Mirsky theorem. The decomposed matrices can be expressed in a block matrix form:
\begin{equation}
    \mathcal{D}(\mathbf{W}) = \underbrace{\mathbf{L}}_{\mathbf{U}\mathbf{\Sigma}}
    \underbrace{
        \begin{bmatrix}
            \mathbf{R}_1,\:
            \mathbf{R}_2,\:
            \cdots,\:
            \mathbf{R}_g,
        \end{bmatrix}
    }_{\mathbf{V}^\top}
\end{equation}
where $\mathbf{L}\!=\!\mathbf{U}\mathbf{\Sigma}$ and $\mathbf{R}_i$ is the $i$-th submatrix of $\mathbf{V}^\top$ which is partitioned into $g$ groups. 

Following the distributive property of block matrices, $\mathbf{L}$ is multiplied with all $\mathbf{R}_i$ matrices, approximating $\mathbf{W}_i$ (i.e., $\mathbf{W_i} \approx \mathbf{L}\mathbf{R}_i$). However, according to the Eckart–Young–Mirsky theorem, we know that $\mathbf{L}\mathbf{R}_i$ is not necessarily the optimal approximation of $\mathbf{W}_i$ since $\mathbf{L}\mathbf{R}_i$ may not be the SVD of $\mathbf{W}_i$. Hence, 
\begin{equation}
    ||\mathbf{W}_i - \mathcal{D}(\mathbf{W}_i)||_\mathrm{F}
    \leq
    ||\mathbf{W}_i - \mathbf{L}\mathbf{R}_i||_\mathrm{F}  \quad\quad \forall i
    \label{eqn:4}
\end{equation}
where $\mathcal{D}(\mathbf{W}_i)$ is the truncated SVD of $\mathbf{W}_i$. Note that RHS represents the reconstruction error of $\mathbf{W}_i$ of the group low-rank compression method, and LHS represents that of the traditional method. 

The Eq. (\ref{eqn:4}) implies that the inequality should hold also for the summation over all $i$ of the squares of the norms:
\begin{equation}
    \underbrace{\sum_{i = 1}^{g} ||\mathbf{W}_i - \mathcal{D}(\mathbf{W}_i)||_\mathrm{F}^2}_{\varepsilon_g^2}
    \leq
    \underbrace{\sum_{i = 1}^{g} ||\mathbf{W}_i - \mathbf{L}\mathbf{R}_i||_\mathrm{F}^2 }_{\varepsilon^2}
\end{equation}
where LHS is the square of $\varepsilon_g$ and RHS is the square of $\varepsilon$.
Since the square function is monotonic for non-negative real numbers and the Frobenius norms are also non-negative, the inequality holds even after taking the square root of both sides. Doing so yields Eq. \ref{eqn:theorem1} and concludes the proof. $\hfill\ensuremath{\Box}$

By Theorem 1, the proposed method guarantees a smaller reconstruction error than the traditional low-rank compression method, promising an improved accuracy performance. Although the performance boost comes at a cost of additional $\mathbf{L}_i$ matrices, note that these matrices are mapped to the idle rows. Therefore, in the context of IMC arrays, the proposed group low-rank compression could potentially offer accuracy gains at no cost, if the number of groups is chosen wisely. Nonetheless, the proposed Theorem 1 is significant as it is universally applicable to all matrices and neural network layers such as convolutional layers and linear layers. 


\mysubsection{SDK for Low-Rank Compression.}
To improve on the low column utilization issue, we seek to integrate SDK mapping \cite{vwc-sdk} together with the low-rank decomposition technique. Since the SDK mapping inherently uses more columns than the im2col mapping, its low-rank decomposed version should also utilize more columns for parallel processing. However, the formulation to derive the low-rank decomposition of SDK mapping is non-trivial. To this end, we first propose a rigorous mathematical description of the SDK mapping method and then derive a low-rank decomposition formula with respect to the SDK mapping. 

\vspace{1em}\noindent\textbf{Theorem 2.} \textit{Given a weight matrix $\mathbf{W}$, and its low-rank decomposed matrices, $\mathbf{L}$ and $\mathbf{R}$, low-rank approximation of the SDK mapping of $\mathbf{W}$ is given by:}
\begin{equation}
    \mathcal{D}(\,\operatorname{SDK}(\mathbf{W})\,) = (\mathbf{I}_N \otimes \mathbf{L}) \operatorname{SDK}(\mathbf{R})
    \label{eqn:theorem2}
\end{equation}
\textit{where $\mathbf{I}_N$ is the identity matrix of size $N \times N$, $N$ is the number of parallel outputs in the SDK mapping, $\otimes$ denotes the Kronecker product, and $\operatorname{SDK}(\cdot)$ denotes the SDK operator that generates SDK mapping for a given matrix.}


\begin{figure*}[t]
\centering
\includegraphics[trim={0cm 0cm 0cm 0cm}, clip, width=0.85\linewidth]{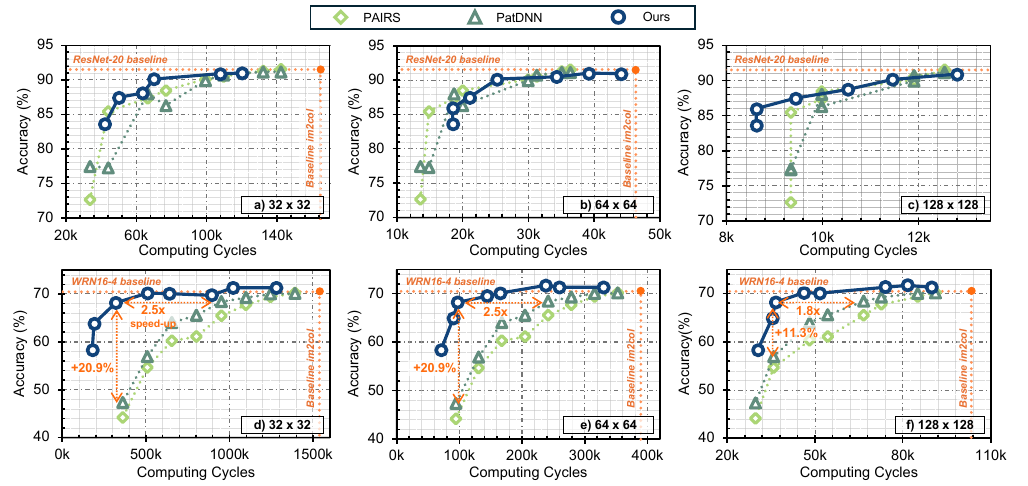}
\vspace{-1em}
\caption{
Accuracy and computing cycle of pattern-pruning methods vs. the proposed method evaluated for ResNet-20 and WRN16-4 with varying array sizes. 
}
\vspace{-1em}
\label{fig:result}
\end{figure*}

\vspace{1em}\noindent\textit{Proof.} A convolutional kernel matricized by im2col mapping method can be described as $\mathbf{W} = [\mathbf{w}_1, \mathbf{w}_2, \cdots, \mathbf{w}_m]^\top$ where $\mathbf{w}_i \in \mathbb{R}^{1 \times n}$ is a vectorized output channel of a convolutional kernel. 
Then the SDK mapping, $\operatorname{SDK}(\mathbf{W}) \in \mathbb{R}^{Nn \times b}$ can be expressed as a linear transformation of $\mathbf{W}$. 
\begin{equation}
    \label{eqn:sdk}
    \begin{split}
        \operatorname{SDK}(\mathbf{W}) & = [\mathbf{P}_1 \mathbf{W}^\top, \mathbf{P}_2 \mathbf{W}^\top, \cdots, \mathbf{P}_N \mathbf{W}^\top]^\top \\
        & = 
        \underbrace{
        \begin{bmatrix}
            \mathbf{W} & \mathbf{0} & \cdots & \mathbf{0} \\
            \mathbf{0} & \mathbf{W} & \cdots & \mathbf{0} \\
            \vdots & \vdots & \ddots & \vdots \\
            \mathbf{0} & \mathbf{0} & \cdots & \mathbf{W} \\
        \end{bmatrix}}_{\mathbf{W}_b \in \mathbb{R}^{Nm \times Nn}}
        \underbrace{
        \begin{bmatrix}
            \mathbf{P}_1^\top \\
            \mathbf{P}_2^\top \\ 
            \vdots \\
            \mathbf{P}_N^\top 
        \end{bmatrix}}_{\mathbf{P} \in \mathbb{R}^{Nn \times b}}
    \end{split}
\end{equation}
where $\mathbf{P}_s \in \mathbb{R}^{b \times n}$ is the $s$-th padding matrix. $N$ is the total number of parallel outputs, which is determined by the PW dimension, and $b$ is the input dimension of the flattened PW. 
The role of the padding matrix is to insert zero column vectors into $\mathbf{W}$, such that the elements of the kernels are appropriately shifted and aligned with the PW input. $\mathbf{P}_s$ can be built from a square identity matrix followed by the insertion of zero row vectors in a specific pattern that is dictated by the SDK mapping. Then the element of $\mathbf{P}_s$ at index $i,j$ is defined as:
\begin{equation}
    [\mathbf{P}_s]_{i,j} =
    \begin{cases}
        1 \quad \text{if } \, i = f(j) \\
        0 \quad \text{otherwise}
    \end{cases}
\end{equation}
where $f(\cdot)$ is a mapping function that describes the insertion locations of the zero column vectors. 

Now we can substitute low-rank compressed matrix of the im2col mapping, $\mathbf{W} = \mathbf{LR}$, in to equation (\ref{eqn:sdk}).
\begin{equation}
    \operatorname{SDK}(\mathbf{W}) = [\mathbf{P}_1 \mathbf{R}^\top \mathbf{L}^\top, \mathbf{P}_2 \mathbf{R}^\top \mathbf{L}^\top, \cdots, \mathbf{P}_N \mathbf{R}^\top \mathbf{L}^\top]^\top
\end{equation}
Then, instead of factoring out the entire $\mathbf{LR}$, which would give us the equivalent formulation as in  (\ref{eqn:sdk}), we can solely factor out $\mathbf{L}$ in the form of block diagonal matrix: 
\begin{equation}
    \operatorname{SDK}(\mathbf{W}) = 
    \underbrace{
    \begin{bmatrix}
        \mathbf{L} & \mathbf{0} & \cdots & \mathbf{0} \\
        \mathbf{0} & \mathbf{L} & \cdots & \mathbf{0} \\
        \vdots & \vdots & \ddots & \vdots \\
        \mathbf{0} & \mathbf{0} & \cdots & \mathbf{L} \\
    \end{bmatrix}}_{\tilde{\mathbf{L}} \in \mathbb{R}^{Nm \times Nk}}
    \underbrace{
    \begin{bmatrix}
        \mathbf{R} \mathbf{P}_1^\top \\
        \mathbf{R} \mathbf{P}_2^\top \\ 
        \vdots \\
        \mathbf{R} \mathbf{P}_N^\top 
    \end{bmatrix}}_{\tilde{\mathbf{R}} \in \mathbb{R}^{Nk \times b}}
    \label{eqn:10}
\end{equation}
where the $\tilde{\mathbf{L}}$ and $\tilde{\mathbf{R}}$ are the low-rank decomposed matrices of the SDK mapping. We can see that $\tilde{\mathbf{L}}$ can be compactly denoted as $\mathbf{I}_N \otimes \mathbf{L}$ and $\tilde{\mathbf{R}} = \operatorname{SDK} (\mathbf{R})$. Plugging in the two expressions for $\tilde{\mathbf{L}}$ and $\tilde{\mathbf{R}}$ into Eq. (\ref{eqn:10}) yields Eq. (\ref{eqn:theorem2}) and concludes the proof. $\hfill\ensuremath{\Box}$

The proposed method is graphically illustrated in Fig. \ref{fig:proposed}b.

\section{Experiments and Results}
\mysubsectionNospace{Experimental Setup.}
To evaluate and demonstrate the effectiveness of the proposed method, we employed ResNet-20 and Wide-ResNet16-4 (WRN16-4) for image classification tasks on CIFAR-10 and CIFAR-100 datasets, respectively. 
Here, ResNet-20 was trained with expansion parameter set to 1 (i.e., the first basic block has 16 input/output channels). 
Weights and activations of all deep learning models were both quantized 4 bit, and the models were trained following the quantization aware training framework proposed in \cite{anyprecision}. 
We did not compress the very first convolution layer and the last linear layer, as they are known to be highly sensitive to perturbations and are often processed on digital computing units that support floating point operations \cite{vwc-sdk, sdk}. 
Proposed low-rank compressed models were trained from scratch for 250 epochs, where as the pattern-pruned counterparts were fine-tuned for 20 epochs from a pre-trained model. The pre-trained model was trained for 200 epochs. We experimented for three trials using different seeds.

\begin{figure}
    \centering
    \includegraphics[width=0.85\columnwidth]{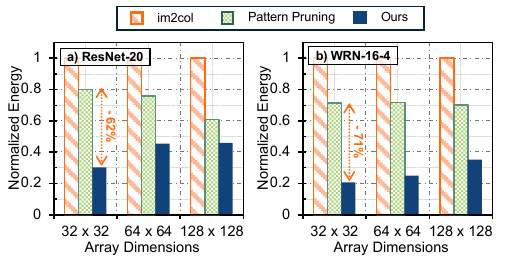}
    \vspace{-1em}
    \caption{Energy consumption of the pattern-pruning methods vs. the proposed methods evaluated for ResNet-20 and WRN16-4 with varying array sizes.}
    \vspace{-1em}
    \label{fig:hw-perf}
\end{figure}

\mysubsection{Comparisons with pattern pruning methods.} 
Table \ref{tab:ours} presents the model accuracy and computing cycle for a low-rank compressed model for different combinations of group and rank. 
The rank of each layer was configured uniformly to the number of output channels, $m$, divided by a constant factor, in this case, 2, 4, 8, and 16. 
Also, the number of groups is set to either 1, 2, 4, or 8. 
Fig. \ref{fig:result} presents a comprehensive overview and comparisons of the proposed low-rank compression method versus the existing pattern-pruning approaches. Baseline accuracies and computing cycles of unpruned models are presented by the orange dotted line. The first row of figures presents experiments conducted on ResNet-20, whereas the second row shows the results for WRN16-4. For pattern-pruning baselines, we plotted the results for entries ranging from 1 to 8, whereas, for our proposed method, we selectively plotted the combinations of rank and group that form the Pareto front for conciseness and clarity. The result demonstrates the effectiveness of the proposed compression method, achieving on-par performance with pattern-pruning approaches on ResNet-20, and significantly outperforming them on WRN16-4. From Fig. \ref{fig:result}d, we can see that our proposed approach can achieve up 2.5$\times$ speedup and +20.9\% accuracy boost compared to the pruning counterparts. 

To evaluate and compare the hardware performance, we have built a simulator based on NeuroSIM~\cite{bib:neurosim}  and ConvMapSIM~\cite{bib:convmapsim} that measures the energy consumption of the proposed and pattern pruning methods. We measured the energy consumption for both ResNet-20 and WRN16-4 networks for varying array sizes. Fig. \ref{fig:hw-perf} plots the normalized energy consumption of the two compression methods against im2col method. For low-rank compressed models, we employ the model with group = 4 and rank = $m/8$, which exhibits high accuracy (less than 1 or 2\% drop from the uncompressed model) while achieving significant computing cycle reduction. For pattern-pruned models, we employ the model pruned with 6-entries, which achieves almost identical accuracy performance as our low-rank model. The results show that the proposed method is more energy-efficient than the pattern-pruned models for both networks across all array dimensions. For smaller arrays, the proposed method could improve energy saving by up to 71\% when compared against the pattern-pruning method and up to 80\% against im2col method. 

We highlight once again, that unlike pruning approaches that necessitate additional peripheral circuitry to combat misalignment and dislocation issues, the proposed method can be adopted on any IMC array and is free of such overheads, and yet achieves better performance in both accuracy and computing cycles. This result is significant and underscores the potential impact of the proposed method when integrated with various deep learning networks and IMC architectures. 

\begin{figure}
    \centering
    \includegraphics[width=0.85\columnwidth]{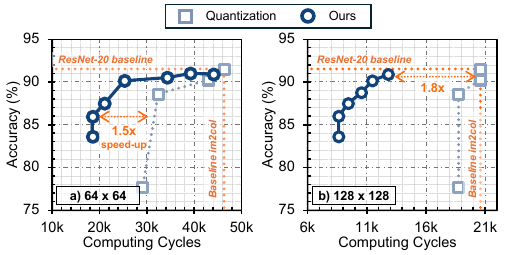}
    \vspace{-1em}
    \caption{Comparison of accuracy and computing cycle performance between low-rank compression models and quantized models.}
    \vspace{-1em}
    \label{fig:vs_quant}
\end{figure}

\begin{figure}
    \centering
    \includegraphics[width=0.85\columnwidth]{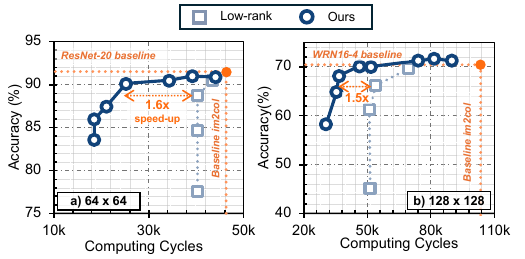}
    \vspace{-1em}
    \caption{Comparison of accuracy and computing cycle performance between low-rank compression models with and without the proposed techniques.}
    \vspace{-1em}
    \label{fig:ablation}
\end{figure}

\mysubsection{Comparisons with quantization methods.} To enrich our evaluation, we also compare our proposed method against quantization with varying bit precision. 
We trained dedicated 1, 2, 3, and 4 bit quantized models of ResNet-20 using a QAT framework and a DoReFa quantizer.
The accuracies and computing cycles of the quantized models for array dimensions of 64$\times$64 and 128$\times$128 are plotted in Fig. \ref{fig:vs_quant}. It can be seen that the proposed low-rank compression method outperforms quantized models, achieving up to 1.8$\times$ speed-up. 

\mysubsection{Comparisons with traditional low-rank compression.} 
As shown in Fig. \ref{fig:ablation} and Table \ref{tab:ours}, the proposed method consistently outperforms the traditional low-rank compressed baseline models (where the proposed SDK mapping and group low-rank compression technique are not applied). 
Whereas the prior low-rank method can reduce the computing cycles to 54K and 40K in the WRN16-4 and ResNet-20 networks, respectively, the proposed method significantly reduces them to 37K in WRN16-4 and 25K in ResNet-20. This is equivalent to 1.5$\times$ and 1.6$\times$ speedup in WRN16-4 and ResNet-20, respectively, due to better array utilization with SDK mapping. The gain is more notable on larger arrays, where SDK mapping can be better explored for more parallel computation.  
On the other hand, the proposed method also boasts significant boosts in accuracy even at lower values of rank, thanks to the use of group low-rank compression. It can be seen from Table \ref{tab:ours}, that with the increasing number of groups, even with just 2, we witness significant mitigation of accuracy drop. 

\begin{table}[]
    \centering
    \caption{Results on low-rank compression}
    \vspace{-1em}
    \includegraphics[width=0.9\columnwidth]{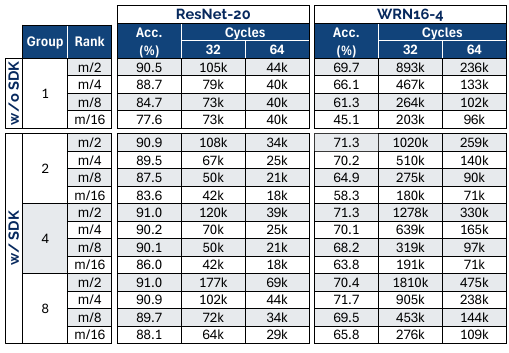}
    \vspace{-2em}
    \label{tab:ours}
\end{table}

\section{Conclusion}
In this study, we tackled the challenge of efficiently compressing models tailored to IMC architectures to enhance computational efficiency without the significant area and energy overheads typical of traditional pruning methods. Our approach introduced low-rank compression techniques integrated with novel SDK and group low-rank convolution strategies, mitigating issues such as suboptimal IMC array utilization and accuracy compromises. Through rigorous experiments on ResNet-20 and WRN16-4 using CIFAR-10 and CIFAR-100 datasets, our method demonstrated its potential by matching or surpassing the performance of existing pruning techniques while significantly reducing computational cycles. This research not only offers a viable alternative to conventional pruning but also opens new avenues for optimizing deep neural networks for IMC architectures, offering 
paving the way for their more efficient deployment in real-world applications.

\section*{Acknowledgement}
This work was partly supported by
the National Research Foundation of Korea (NRF) grant (No. RS-2024-00345732); 
the Institute for Information \& communications Technology Planning \& Evaluation (IITP) grants (RS-2020-II201821, IITP-2021-0-02052, RS-2019-II190421, RS-2021-II212068); 
the Technology Innovation Program (RS-2023-00235718, 23040-15FC) funded by the Ministry of Trade, Industry \& Energy (MOTIE, Korea) (1415187505); 
Samsung Electronics Co., Ltd (IO230404-05747-01); 
and the BK21-FOUR Project.


\bibliographystyle{IEEEtran}
\bibliography{IEEEabrv,mybibfile}

\begin{thebibliography}{10}
\providecommand{\url}[1]{#1}
\csname url@samestyle\endcsname
\providecommand{\newblock}{\relax}
\providecommand{\bibinfo}[2]{#2}
\providecommand{\BIBentrySTDinterwordspacing}{\spaceskip=0pt\relax}
\providecommand{\BIBentryALTinterwordstretchfactor}{4}
\providecommand{\BIBentryALTinterwordspacing}{\spaceskip=\fontdimen2\font plus
\BIBentryALTinterwordstretchfactor\fontdimen3\font minus \fontdimen4\font\relax}
\providecommand{\BIBforeignlanguage}[2]{{%
\expandafter\ifx\csname l@#1\endcsname\relax
\typeout{** WARNING: IEEEtran.bst: No hyphenation pattern has been}%
\typeout{** loaded for the language `#1'. Using the pattern for}%
\typeout{** the default language instead.}%
\else
\language=\csname l@#1\endcsname
\fi
#2}}
\providecommand{\BIBdecl}{\relax}
\BIBdecl

\bibitem{pipelayer}
L.~Song, X.~Qian, H.~Li, and Y.~Chen, ``Pipelayer: A pipelined reram-based accelerator for deep learning,'' in \emph{HPCA}, 2017.

\bibitem{im2col}
K.~Yanai \emph{et~al.}, ``Efficient mobile implementation of a {CNN}-based object recognition system,'' in \emph{ACM MM}, 2016.

\bibitem{sdk}
Y.~Zhang \emph{et~al.}, ``Efficient and robust rram-based convolutional weight mapping with shifted and duplicated kernel,'' \emph{TCAD}, 2020.

\bibitem{vwc-sdk}
J.~Rhe \emph{et~al.}, ``Vwc-sdk: Convolutional weight mapping using shifted and duplicated kernel with variable windows and channels,'' \emph{JETCAS}, 2022.

\bibitem{patdnn}
W.~Niu \emph{et~al.}, ``Patdnn: Achieving real-time dnn execution on mobile devices with pattern-based weight pruning,'' in \emph{ASPLOS}, 2020.

\bibitem{pairs}
J.~Rhe \emph{et~al.}, ``Pairs: Pruning-aided row-skipping for sdk-based convolutional weight mapping in processing-in-memory architectures,'' in \emph{ISLPED}, 2023.

\bibitem{exploring}
F.-H. Meng \emph{et~al.}, ``Exploring compute-in-memory architecture granularity for structured pruning of neural networks,'' \emph{JETCAS}, 2022.

\bibitem{unst_pruning}
S.~Han \emph{et~al.}, ``Learning both weights and connections for efficient neural network,'' \emph{NeurIPS}, 2015.

\bibitem{zero-skipping}
J.-H. Kim \emph{et~al.}, ``Z-pim: A sparsity-aware processing-in-memory architecture with fully variable weight bit-precision for energy-efficient deep neural networks,'' \emph{JSSC}, 2021.

\bibitem{islped_row_merging}
K.~E. Jeon, J.~Rhe, H.~Bang, and J.~H. Ko, ``Weight-aware activation mapping for energy-efficient convolution on pim arrays,'' in \emph{2023 IEEE/ACM International Symposium on Low Power Electronics and Design (ISLPED)}, 2023, pp. 1--6.

\bibitem{structured}
J.~Meng \emph{et~al.}, ``Structured pruning of rram crossbars for efficient in-memory computing acceleration of deep neural networks,'' \emph{TCAS-II}, 2021.

\bibitem{pimprune}
C.~Chu \emph{et~al.}, ``Pim-prune: Fine-grain dcnn pruning for crossbar-based process-in-memory architecture,'' in \emph{DAC}, 2020.

\bibitem{flexible}
L.~Zheng \emph{et~al.}, ``A flexible yet efficient dnn pruning approach for crossbar-based processing-in-memory architectures,'' \emph{TCAD}, 2022.

\bibitem{denton_low-rank}
E.~L. Denton \emph{et~al.}, ``Exploiting linear structure within convolutional networks for efficient evaluation,'' \emph{NeurIPS}, vol.~27, 2014.

\bibitem{trp_low-rank}
Y.~Xu \emph{et~al.}, ``{TRP}: Trained rank pruning for efficient deep neural networks,'' \emph{IJCAI}, 2020.

\bibitem{yani_low-rank}
Y.~Ioannou \emph{et~al.}, ``Training cnns with low-rank filters for efficient image classification,'' \emph{ICLR}, 2016.

\bibitem{anyprecision}
H.~Yu \emph{et~al.}, ``Any-precision deep neural networks,'' in \emph{AAAI}, 2021.

\bibitem{bib:neurosim}
X.~Peng \emph{et~al.}, ``Dnn+neurosim v2.0: An end-to-end benchmarking framework for compute-in-memory accelerators for on-chip training,'' \emph{IEEE TCAD}, vol.~40, no.~11, pp. 2306--2319, 2021.

\bibitem{bib:convmapsim}
K.~E. Jeon \emph{et~al.}, ``Convmapsim: Modeling and simulating convolutional weight mapping for pim arrays,'' in \emph{IEEE AICAS}, 2024, pp. 417--421.

\end{thebibliography}

\end{document}